\documentclass[aps,amssymb,floatfix,prd,amsmath,preprintnumbers]{revtex4}
\usepackage{epstopdf}
\usepackage{capt-of}
\usepackage{graphicx}  
\usepackage{dcolumn}   
\usepackage{bm}
\begin{document}
\input epsf.tex
\title{\bf Dynamical features of an anisotropic cosmological model}

\author{B. Mishra$^1$\footnote{Corresponding Author, E-mail: bivudutta@yahoo.com}, Sankarsan Tarai$^1$, S.K. Tripathy$^2$\footnote{E-mail: tripathy\_sunil@rediffmail.com} \\ 
$^1$Department of Mathematics, Birla Institute of Technology and Science-Pilani,\\
Hyderabad Campus, Hyderabad-500078, India\\
$^2$Department of Physics, Indira Gandhi Institute of Technology, \\
Sarang, Dhenkanal, Odisha-759146, India.
}

\affiliation{ }

\begin{abstract}
\begin{center}
\textbf{Abstract}
\end{center}
The dynamical features of Bianchi type $VI_h$ $(BVI_h)$ universe are investigated in $f(R,T)$ theory of gravity. The field equations and the physical properties of the model are derived considering a power law expansion of the universe.  The effect of anisotropy on the dynamics of the universe as well on the energy conditions are studied. The assumed anisotropy of the model is found to have substantial effects on the energy condition and dynamical parameters.
\end{abstract}

\maketitle


\section*{1. Introduction} 

In the last three decades, the standard model of cosmology has gained a lot of research attention. The prime reason behind this is its ability to address complex observational issues with a simple theoretical structure. The two major successes of this model are $(i)$ the explanation of the observed light element abundances in early universe and $(ii)$ the prediction of the relic Cosmic Microwave Background (CMB). Over a period of time, cosmological theories have come up with a number of novel predictions including the possible existence of topological defects, extra dimensions, inflation, relic non baryonic dark matter candidates.

Substantial progress have been made in recent years in acquisition of cosmological data both in quality and quantity. This advancement obviously lowers the gap between theory and observation. Also it provides a window to understand the physics of the very early universe. In the past two decades, the supernovae cosmology project group and the high- z supernovae group have presented enough evidences with observations and theoretical justification that the universe is undergoing an  accelerating expansion at the present epoch \cite{Schmi98, Riess98, Perlm99}. A mysterious energy form called dark energy (DE) may be responsible for such a phenomena. In the frame work of General Relativity (GR), cosmological constant is an obvious choice for dark energy. However,  some modified theories of gravity have been proposed in recent time that provide suitable platform to understand the cosmic speed up phenomena without the need of any dark energy form. Though there have been several theoretical models proposed in literature; cosmologists are yet to reveal the cosmological origin, nature and  behaviour of dark energy. In this context, modified theories of gravity have attracted a lot of attention. Prominent gravity theories with geometry modification include $f(R)$ theory\cite{Nojir07, Trip16}, $f(T)$ gravity \cite{ Myrza11, Harko14a} and $f(R,T)$ gravity \cite{Harko11} where $R$ is the Ricci Scalar and $T$ is the trace of the energy momentum tensor. In these theories, the usual Ricci Scalar $R$ in the gravitational action equation is replaced by a more general function of $R$ or a combined function of $R$ and $T$.

It is interesting to investigate anisotropic models in modified theories. Although there have been a lot of debate going on in the issue of cosmic anisotropy, anisotropic models can be more interesting in the sense that, they are quite general than the usual FRW models and may provide some interesting cosmological results \cite{Saadeh16, SKT14, Zhao13, Ant10, Mar12}.  Shamir \cite{Shami15} , Mishra and Sahoo \cite{Mishr14}, Sharif and Zubair \cite{Shari12, Shari14}, Jamil et al.\cite{Jamil12}  have investigated Bianchi type cosmological models in $f(R,T)$ theory. Moreover, Mishra et al. \cite{Mishr15} have constructed the cosmological model in a quadratic form of $f(R,T)$ gravity where the space time is considered to be a non-static and the matter field is perfect fluid where as Mishra et al. \cite{Mishr16} have studied the dynamics of an anisotropic universe in a Bianchi type $VI_h$ space time. Shamir \cite{Shami15a} has obtained the solutions of  the LRS Bianchi type I space-time with the assumption of a relationship between the metric potentials of the space-time. Yousaf et al. \cite{Yous16a, Yous16b} have studied the irregularity energy density factor in homogeneous universe and also investigated the possibility of evolutionary behaviour of the compact objects. Mishra and Vadrevu \cite{Mishr17} have studied the cylindrically symmetric cosmological model of the universe with a quadratic form of the Ricci scalar. Sharif and Nawazish \cite{Shari17} have investigated some cosmological models for Bianchi type I, III and Kantoskwi-Sachs space-time using Noether symmetry approach. Also, Sharif and Nawazish \cite{Shari17a} have explored the Noether and Noether gauge symmetries of anisotropic cosmological model in $f(R,T)$ gravity. Shabani and Ziaie \cite{Shab17} have studied the existence and stability of Einstein universe in $f(R,T)$ gravity. Zubair et al. have  analysed the stability of cylindrically symmetric objects with anisotropic fluids \cite{Zubair17}. Agrawal and Pawar \cite{Agrawal17a,Agrawal17b} have studied the magnetized domain wall and quark and strange quark matter in $f(R,T)$ gravity. There have been a good number of theoretical investigation on the $f(R,T)$ gravity theory in recent times in cosmology and astrophysics \cite{Sahu17, Sahoo16, Yousaf17, Nou17, Hus17, Moraes17, Harko14}.

The paper is organized as follows: In section 2, the basic formalism of $f(R,T)$ gravity and the consequent field equations have been set up for Bianchi type $VI_h$ space-time. In this section, we have developed a general mathematical scheme for different physical parameters including the equation of state (EoS) parameter and effective cosmological constant. The dynamical features of the model and the energy conditions for a power law form of mean scale factor are obtained in section 3. The results and discussions of the model are presented in section 4. Finally the conclusion of the present work is given in section 5. 

\section*{2. Basic Formalism}

The field equations for $f(R,T)$ gravity with special choice of $f(R,T)=f(R)+f(T)$ are written as \cite{Harko11, Mishr16}, 
\begin{equation} \label{eq:1}
f_R (R) R_{ij}-\frac{1}{2}f(R)g_{ij}+\left(g_{ij}\Box-\nabla_i \nabla_j\right)f_R(R)=\left[8\pi+f_T(T)\right]T_{ij}+\left[pf_T(T)+\frac{1}{2}f(T)\right]g_{ij}
\end{equation}
where $f_R=\frac{\partial f(R)}{\partial R}$ and $f_T=\frac{\partial f(T)}{\partial T}$ are the partial differentiations of $f(R,T)$ with respect to their respective variables. Here $p$ is the pressure of the cosmic fluid distribution. If we consider $f(R)$ and $f(T)$ in linear form respectively as $f(R)=\lambda R $ and $f(T)=\lambda T$, $\lambda$ being a constant scaling factor, eq. \eqref{eq:1} can be reduced to   
\begin{equation} \label{eq:2}
R_{ij}-\frac{1}{2}Rg_{ij}=\left(\frac{8\pi+\lambda}{\lambda}\right)T_{ij} + \Lambda (T)g_{ij}
\end{equation}
where $\Lambda(T)=p+\frac{1}{2}T$ is an effective cosmological constant that depends on time . It picks up its evolutionary behaviour through the matter fields. It is worth to mention here that, the scaling factor $\lambda$ can not be zero as the model diverges for this value. Also, it is certain that, one can not recover the corresponding field equations of GR by putting a value of $\lambda$ by hand. However, as it can be seen from our discussion, we may obtain viable models by rescaling the GR equations through this parameter $\lambda$. An important feature of this model is that, the field equation appears to have the same form as that of GR with a time varying cosmological constant and a redefined Einstein constant($\kappa=\frac{8\pi G}{c^4}$, $G$ and $c$ are respectively the Newtonian gravitational constant and speed of light in vacuum). We assume the energy momentum tensor as 

\begin{equation}  \label{eq:3}
T_{ij}=(p+\rho)u_iu_j - pg_{ij}-\rho_B x_ix_j
\end{equation}
where $u^{i}u_{i}=-x^{i}x_{i}=1$ and $u^{i}x_{i}=0$. In a co-moving coordinate system, $u^{i}$ is the four velocity vector. $x^{i}$ represents the direction of anisotropic fluid (here x-direction) and is orthogonal to $u^{i}$. $\rho$ is the energy density and is composed of energy density due to the perfect fluid and anisotropic fluid $\rho_B$. The field equations \eqref{eq:2} for Bianchi type $VI_h$ space-time 

\begin{equation} \label{eq:4}
ds^2 = dt^2 - A^2dx^2- B^2e^{2x}dy^2 - C^2e^{2hx}dz^2
\end{equation}
reduce to
\begin{equation} \label{eq:5}
\frac{\ddot{B}}{B}+\frac{\ddot{C}}{C}+\frac{\dot{B}\dot{C}}{BC}- \frac{h}{A^2}= -\alpha(p-\rho_B) +\frac{\rho}{2}   
\end{equation}
\begin{equation} \label{eq:6}
\frac{\ddot{A}}{A}+\frac{\ddot{C}}{C}+\frac{\dot{A}\dot{C}}{AC}- \frac{h^2}{A^2}=-\alpha p+\frac{1}{2}\left(\rho+\rho_B\right) 
\end{equation}
\begin{equation} \label{eq:7}
\frac{\ddot{A}}{A}+\frac{\ddot{B}}{B}+\frac{\dot{A}\dot{B}}{AB}- \frac{1}{A^2}=-\alpha p+\frac{1}{2}\left(\rho+\rho_B\right) 
\end{equation}
\begin{equation} \label{eq:8}
\frac{\dot{A}\dot{B}}{AB}+\frac{\dot{B}\dot{C}}{BC}+\frac{\dot{C}\dot{A}}{CA}-\frac{1+h+h^2}{A^2}=
\alpha \rho -\frac{1}{2}\left(p-\rho_B\right)   
\end{equation}
\begin{equation} \label{eq:9}
\frac{\dot{B}}{B}+ h\frac{\dot{C}}{C}- (1+h)\frac{\dot{A}}{A}=0.
\end{equation} 

Here $A=A(t),B=B(t), C=C(t)$, $\alpha=\frac{16\pi+3\lambda}{2\lambda}$. In the above field equations, an overhead dot denotes  time derivative. The exponent $h$ is a constant which may take integral values $-1,0,1$. These values decide the behaviour of the model. In some recent papers, Tripathy et al. \cite{SKTem15, SKTem16} have calculated the energy and momentum of anisotropic $BVI_h$ universes and  have shown that the energy and momentum of such universes vanish for $h=-1$. If we assume that, the metric should envisages an isolated universe with null total energy and momentum, then the choice $h=-1$ is preferable to any other value of the exponent. In view of this, we assume this value of $h$ and study the dynamics of the anisotropic universe in presence of anisotropic energy sources. The directional Hubble rates may be considered as $H_x=\frac{\dot{A}}{A}$, $H_y=\frac{\dot{B}}{B}$ and $H_z=\frac{\dot{C}}{C}$. With $h=-1$, it is straightforward to get $H_y=H_Z$ from \eqref{eq:9}. The mean Hubble parameter becomes, $H=\frac{\dot{a}}{a}=\frac{1}{3}(H_x+2H_z)$ where $a$ is the mean scale factor of universe. The set of field equations can be reduced to

\begin{eqnarray}
6(k+2)\dot{H}+27H^2+(k+2)^2 a^{-\left(\frac{6k}{k+2}\right)} &=& (k+2)^2\left[-\alpha(p-\rho_B) +\frac{\rho}{2} \right], \label{eq:10}\\
3(k^2+3k+2)\dot{H}+9(k^2+k+1)H^2-(k+2)^2 a^{-\left(\frac{6k}{k+2}\right)} &=& (k+2)^2\left[-\alpha p+\frac{1}{2}\left(\rho+\rho_B\right)\right],\label{eq:11}\\ 
9(2k+1)H^2-(k+2)^2 a^{-\left(\frac{6k}{k+2}\right)} &=& (k+2)^2\left[\alpha \rho -\frac{1}{2}\left(p-\rho_B\right)\right].\label{eq:12}
\end{eqnarray}

Here, a linear anisotropic relation among the directional Hubble rates is assumed i.e. $H_x=kH_z$. Algebraic manipulation of the field equations \eqref{eq:10}-\eqref{eq:12} yields
\begin{eqnarray}
p &=& -\frac{2}{1-4\alpha^2}[s_1(a)+s_3(a)-(1+2\alpha) s_2(a)],\label{eq:13}\\
\rho &=& \frac{2}{1-4\alpha^2}[s_1(a)-2\alpha s_3(a)],\label{eq:14}\\
\rho_B &=& \frac{2}{1-2\alpha}[s_2(a)-s_1(a)].\label{eq:15}
\end{eqnarray}
Here, $s_1(a)=\frac{1}{(k+2)^2}[6(k+2)\dot{H}+27H^2+(k+2)^2 a^{-\left(\frac{6k}{k+2}\right)}]$, $s_2(a)=\frac{1}{(k+2)^2}[3(k^2+3k+2)\dot{H}+9(k^2+k+1)H^2-(k+2)^2 a^{-\left(\frac{6k}{k+2}\right)}]$ and $s_3(a)=\frac{1}{(k+2)^2}[9(2k+1)H^2-(k+2)^2 a^{-\left(\frac{6k}{k+2}\right)}]$. 

The equation of state (EoS) parameter $\omega=\frac{p}{\rho}$ and the effective cosmological constant $\Lambda$ are obtained as 

\begin{eqnarray}
\omega &=&-1+(1+2\alpha)\left[\frac{s_2(a)-s_3(a)}{s_1(a)-2\alpha s_3(a)}\right], \label{eq:16} \\
\Lambda &=& \frac{1}{(1+2\alpha)}[s_1(a)+s_3(a)].\label{eq:17}
\end{eqnarray}

The dynamical features of the model are decided by the physical quantities given in eqns \eqref{eq:13}-\eqref{eq:17}. However these quantities depend on the mean scale factor. In view of this, we can study a background cosmology and associated dynamics if we presume the behaviour of the mean scale factor. 

\section*{3. Dynamical Features and Energy conditions}

In the study of cosmic dynamics, power law cosmology has gained a lot of attention. Power law cosmology has emerged as an alternative to $\Lambda$CDM model  and considers the evolution of classical fields that are coupled to the curvature of the space in such a way that their contribution to the energy density self-adjusts to cancel the vacuum energy \cite{Doglov97}. The motivation for such a scenario comes from the fact that it does not encounter flatness and the horizon problem at all. Another interesting feature of these models is that they easily accommodate high redshift objects and hence alleviate the age problem. These models are also purged of the fine tuning problem \cite{Doglov82, Ford87}. In power law expansion cosmology, the scale factor is assumed to grow as some power function of cosmic time i.e. $a=t^{m/3}$, where $m$ is a positive constant and can be constrained from observational data on the deceleration parameter (DP) or the jerk parameter (JP). The geometrical features such as Hubble parameter, DP and JP respectively  become $H=\frac{m}{3t}$, $q=-1+\frac{3}{m}$ and $j=\frac{9}{m}\left(\frac{2}{m}-1\right)+1$. In the present work, we have employed such a scale factor to investigate the background cosmology in the presumed modified gravity model. With such an assumption we obtain the pressure, energy density and density of anisotropic fluid source from eqs. (13)-(15) as

\begin{eqnarray}
p &=& -\frac{2}{(1-4\alpha^2)}\left[\frac{\phi_1+2\alpha\phi_2}{(k+2)^2}\right]\frac{1}{t^2}-\frac{2}{(1-2\alpha)}\frac{1}{t^{\frac{2km}{k+2}}},\label{eq:18}\\
\rho &=& \frac{2}{(1-4\alpha^2)}\left[\frac{\phi_3-2\alpha(2k+1)m^2}{(k+2)^2}\right]\frac{1}{t^2}+\frac{2}{(1-2\alpha)}\frac{1}{t^{\frac{2km}{k+2}}}, \label{eq:19}\\
\rho_B &=& \frac{2}{(1-2\alpha)}\left[\frac{\phi_4}{(k+2)^2}\right]\frac{1}{t^2}+\frac{4}{(1-2\alpha)}\frac{1}{t^{\frac{2km}{k+2}}}.\label{eq:20}
\end{eqnarray}
where $\phi_1= (k^2+k-2)m-(k^2-k-3)m^2$, $\phi_2= (k^2+3k+2)m-(k^2+k+1)m^2$, $\phi_3= 3m^2-2(k+2)m$  and $\phi_4= (m^2-m)(2-k-k^2)$ are redefined constants that depend on the choice of the parameters $m$ and $k$.

The equation of state parameter(EoS) $\omega$   and the effective cosmological constant are obtained from \eqref{eq:16} and  \eqref{eq:17} as

\begin{eqnarray}
\omega &=& -1+(1+2\alpha)\left[\frac{(k^2-k)m^2-(k^2+3k+2)m}{(3-2\alpha(2k+1))m^2-(k+2)2m+(1+2\alpha)(k+2)^2 t^{\frac{2(k+2-km)}{k+2}}}\right], \label{eq:21}\\
\Lambda &=& \frac{2}{(1+2\alpha)}\left[\frac{m(m-1)}{k+2}\right]\frac{1}{t^2}.\label{eq:22}
\end{eqnarray}

Energy conditions put some additional constraints on the models \cite{Carroll/2004, Sharif/2013, Moraes17a}. For a perfect fluid distribution the energy conditions are 
\begin{eqnarray}
\text{Null Energy Condition} \textbf{(NEC)}&:& \rho+p\geq 0,\nonumber\\
\text{Weak Energy Condition} \textbf{(WEC)} &:& \rho+p\geq 0, ~~~~~\rho \geq 0,\nonumber\\
\text{Strong Energy Condition} \textbf{(SEC)} &:& \rho+3p\geq 0, ~~~\rho+p\geq 0,\nonumber\\
\text{Dominant Energy Condition} \textbf{(DEC)} &:& \rho \pm p\geq 0,~~~~~ \rho\geq 0 \nonumber
\end{eqnarray}

From \eqref{eq:13}-\eqref{eq:14}, the energy conditions are obtained as 
\begin{eqnarray}
\textbf{NEC}&:& \rho+p=\frac{2[s_2(a)-s_3(a)]}{1-2\alpha}\geq 0,\nonumber\\
\textbf{WEC}&:& \rho =\frac{2}{1-2\alpha}[s_2(a)-s_1(a)]\geq 0,\nonumber\\
\textbf{SEC}&:& \rho+3p=-\frac{2}{1-4\alpha^2}\left[s_1(a)+(3+2\alpha)s_3(a)-3(1+2\alpha)s_2(a)\right]\geq 0, \nonumber\\
\textbf{DEC} &:& \rho -p= \frac{2}{1-4\alpha^2}\left[2s_1(a)-(1+2\alpha)s_2(a)+(1-2\alpha)s_3(a)\right]\geq 0. \nonumber
\end{eqnarray}
NEC is implied in all other energy conditions. Since in our calculation, we assume small values for the scaling constant $\lambda$, the factor $1-2\alpha$ is a negative quantity. Consequently, in order to satisfy the null energy condition and weak energy condition, we require $s_1(a) > s_2(a)$ and $s_3(a)> s_2(a)$. Dominant energy condition requires the inequality $s_2(a) > \frac{2s_1(a)+(1-2\alpha)s_3(a)}{1+2\alpha}$ to be satisfied. For \textbf{SEC}, we need an extra condition $s_1(a)>6\alpha s_2(a)$. We have calculated the energy conditions within our formalism with a power law scale factor. These conditions can be expressed as

\begin{eqnarray}
\textbf{NEC} &:& \rho+p =\frac{2}{1-2\alpha}\left[\frac{k(k-1)m^2-(k+1)(k+2)m}{(k+2)^2}\right]\frac{1}{t^2},\nonumber\\
\textbf{SEC} &:& \rho+3p =\frac{2}{1-2\alpha}\left[\frac{3(k^2-k-2)m^2-(3k-1)(k+2)m}{(k+2)^2}\right]\frac{1}{t^2}\nonumber\\
			&& ~~~~~~~~~~+ \frac{4\alpha}{1-2\alpha}\left[\frac{(3k^2-k+1)m^2-3(k+1)(k+2)m}{(k+2)^2}\right]\frac{1}{t^2}\nonumber\\
			&&~~~~~~~~~~-\frac{4(1+\alpha)}{1-4\alpha^2}\frac{1}{t^{\frac{2km}{k+2}}},\nonumber\\
\textbf{DEC} &:& \rho-p= \frac{2}{1-4\alpha^2}\left[\frac{\left(k^2+3k+8)-4\alpha(k^2+k+1)\right)m^2+2(k^2+k-2)m}{(k+2)^2}\right]\frac{1}{t^2}\nonumber\\
&&~~~~~~~~~ + \frac{4}{1-2\alpha}\frac{1}{t^{\frac{2km}{k+2}}}. \nonumber
\end{eqnarray}

\begin{figure}[ht!]
\begin{center}
\includegraphics[width=0.85\textwidth]{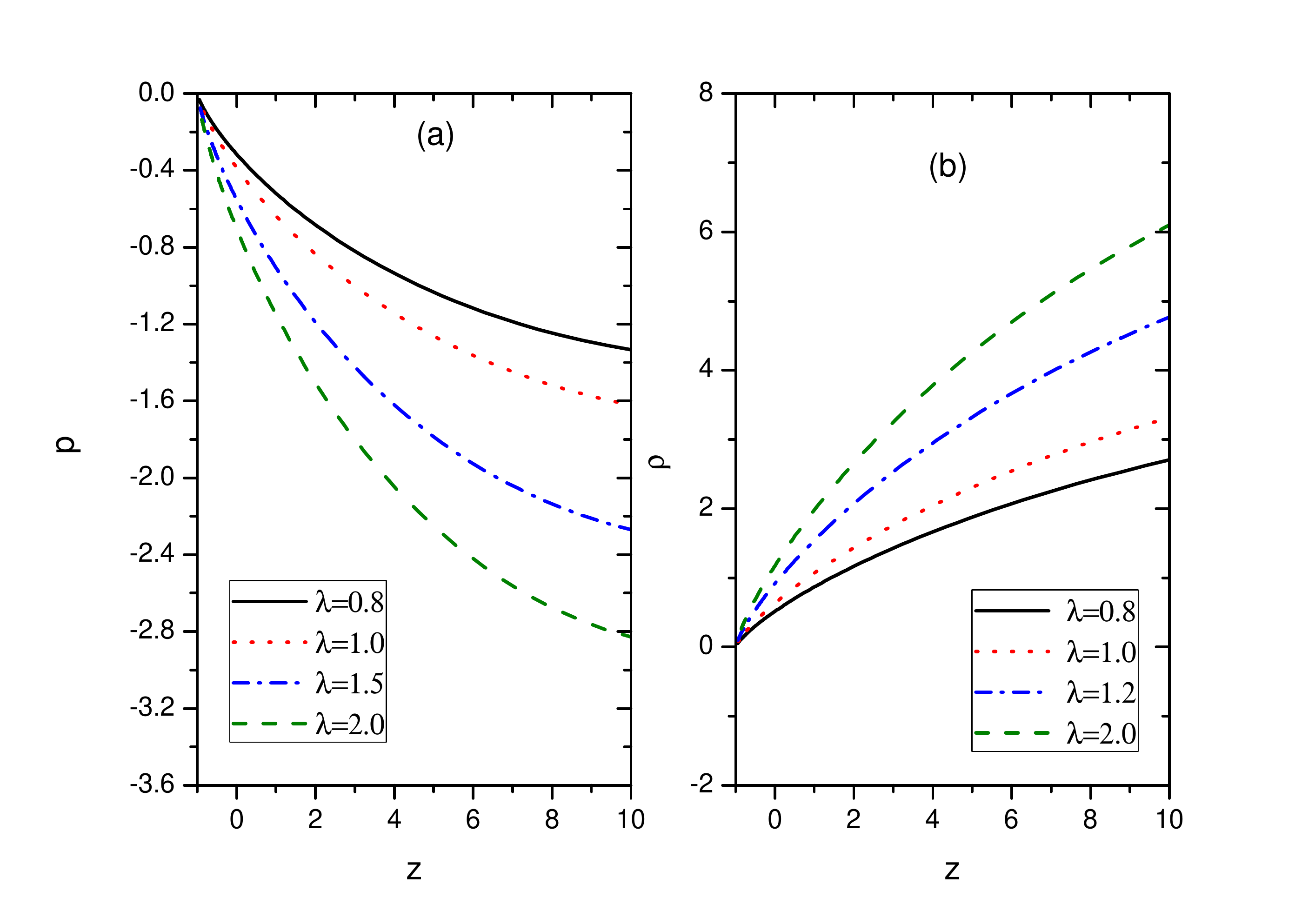}
\caption{(a)Pressure as a function of redshift (b) Energy density as a function of redshift. Four different values of the scaling factor have been considered to assess its effect. Anisotropy parameter is assumed to be $k=0.65$.}
\end{center}
\end{figure}

\section*{4. Results and Discussion}

In Figures 1(a) and (b), we have shown respectively the behaviour of pressure and energy density as function of redshift $z=\frac{1}{a}-1$. The radius scale factor at the present epoch is considered to be 1. The behaviour of the physical quantities are shown for four different values of the scaling constant $\lambda$. The anisotropic parameter $k$ can be considered as a free parameter in the work. However, basing upon the results of our previous work \cite{Mishr16}, we assume a representative value of anisotropy i.e $k=0.65$ for plotting the figures. The exponent $m$ has been constrained from the value of deceleration parameter, $q=-0.598$, obtained from an analysis of observational data \cite{Montiel14}. In general, pressure is obtained to be negative in the whole range of redshift considered in the work. In fact, for a given value of $\lambda$, pressure increases from some large negative value at an early epoch to vanishingly small values at late times. The choice of the scaling parameter $\lambda$ substantially affects the magnitude of pressure. At a given redshift, pressure assumes large negative values with an increase in  $\lambda$. On the other hand, energy density  remains in the positive domain and decreases to small values at late times ( refer to Fig.1(b)). At a given redshift, energy density increases with an increase in $\lambda$.\\

The evolutionary behaviour of the energy density of the anisotropic fluid source, $\rho_B$, is shown in Figure 2. Its evolutionary behaviour is studied by assuming a fixed anisotropy and an accelerating universe with $q=-0.598$. The magnitude of $\rho_B$ is more at an initial epoch compared to late phase. At a late cosmic phase, $\rho_B$ becomes negligible. The variation of $\rho_B$ with different choices of $\lambda$ is also shown in the figure for four representative values of $\lambda$. One can note that, $\rho_B$ increases with an increase in $\lambda$ at a given redshift. The scaling constant also decides the rate of decrement in the energy density of the anisotropic fluid source. Higher the value of $\lambda$, more is the rate of decrement ( or slope) for $\rho_B$. \\

\begin{figure}[ht!]
\begin{center}
\includegraphics[width=0.85\textwidth]{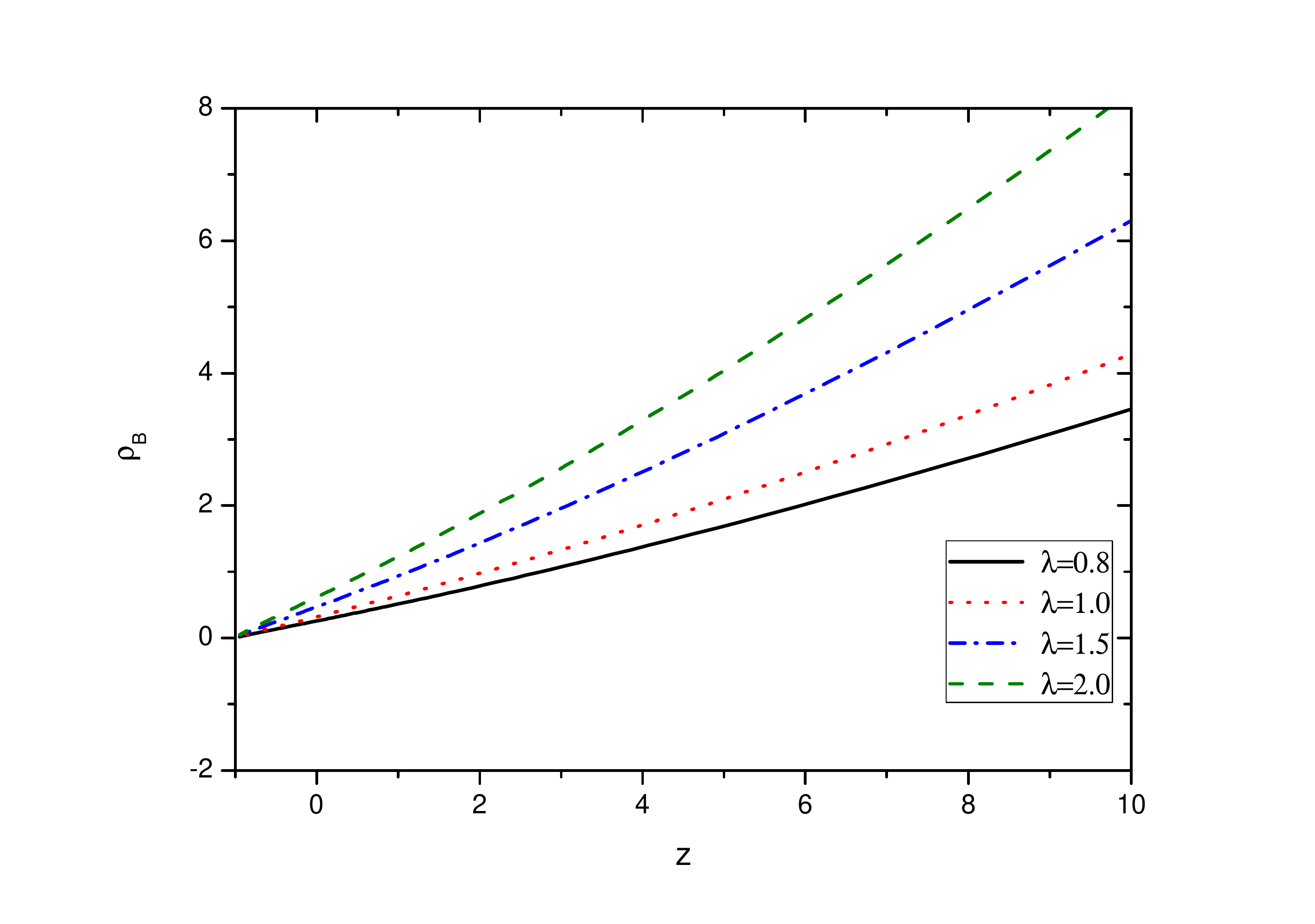}
\caption{Evolution of energy density of the anisotropic fluid source for four different values of the scaling constant $\lambda$.}
\end{center}
\end{figure}


It is evident from eq.(21) that the EoS parameter $\omega$ evolves with time that depends on three parameters $\lambda, m$ and $k$. In Figure 3(a), the time evolution of $\omega$ is shown for four different values of $\lambda$. The anisotropy parameter is considered to be $k=0.65$ and $m$ is constrained from the deceleration parameter. $\omega$ remains in the negative domain  favouring a quintessence phase. EoS parameter decreases with  time. The choice of $\lambda$ value affects the equation of state, mostly it affects the slope of the evolution curve. The slope increases for higher value of $\lambda$. $\omega$ evolves from $\omega=-1+(1+2\alpha)\left[\frac{(k^2-k)m-(k^2+3k+2)}{(3-2\alpha(2k+1))m-2(k+2)}\right]$ at the beginning to some higher negative values compared to this one at late phase of evolution. The value of $\omega$ at late times depends on the choice of $\lambda$. For the given anisotropic parameter i.e. $k=0.65$, the value of $\omega$ in the present epoch becomes $-0.614, -0.611, -0.604$ and $-0.597$ corresponding to $\lambda=0.8, 1.0, 1.5$ and 2.0 respectively.\\

The effective cosmological constant $\Lambda$ decreases quadratically with time and depends on the parameters $\lambda, m$ and $k$. In Figure 3(b), $\Lambda$ is shown as a function of redshift. It is a positive quantity for all the choices of $\lambda$. We have presented the modified gravity model in such a manner that, the theory behaves as that of GR with a time varying cosmological constant. In case of GR, the late time cosmic speed up phenomena, requires a positive non zero cosmological constant that should dynamically roll down to a value close to zero at late times. From our model, we obtain a similar behaviour of the effective cosmological constant. The evolutionary behaviour is more or less the same as that of the energy density of the anisotropic fluid $\rho_B$.\\

\begin{figure}[ht!]
\begin{center}
\includegraphics[width=0.85\textwidth]{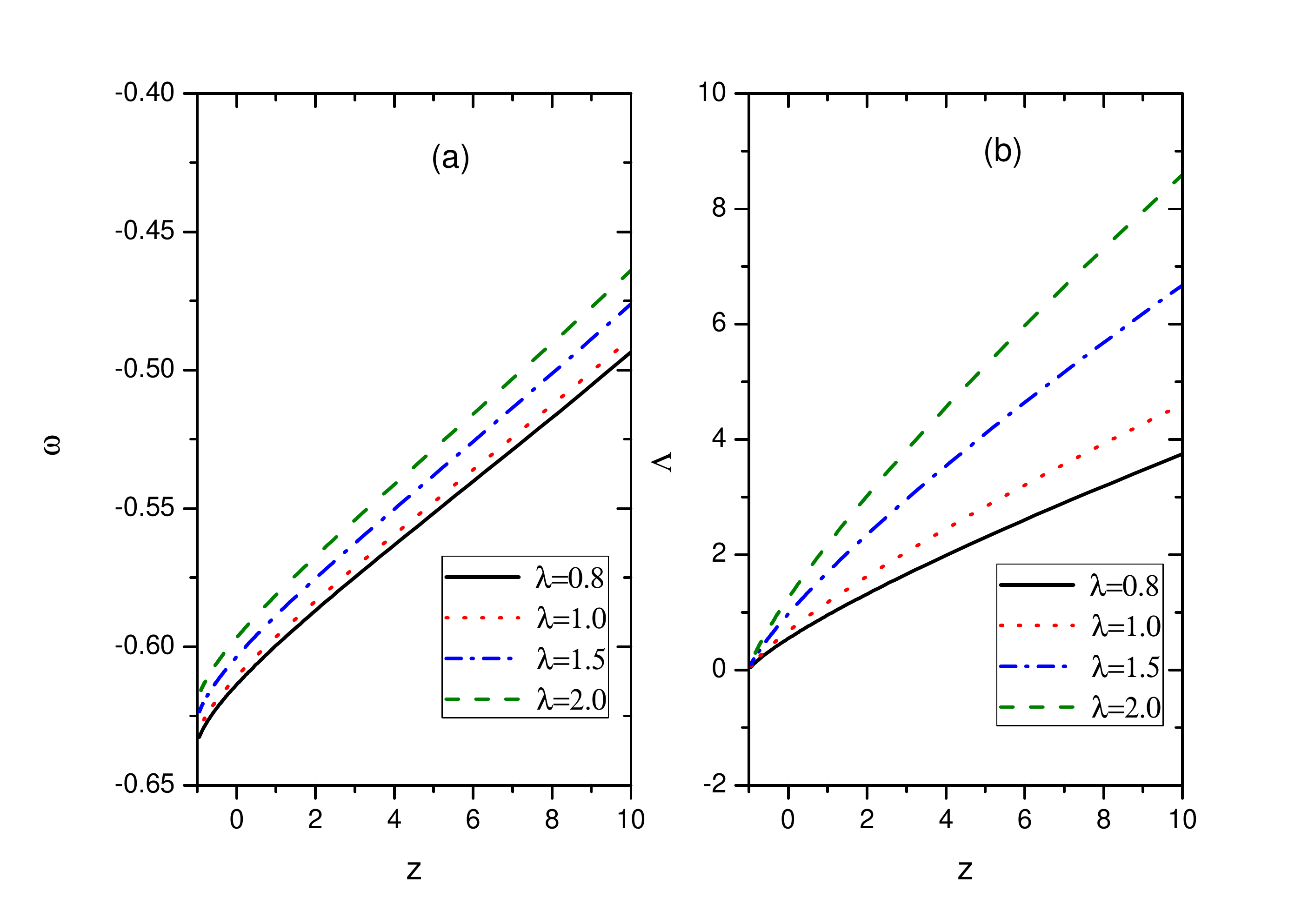}
\caption{(a) Evolution of the EoS parameter. (b) Effective cosmological constant as a function of redshift.}
\end{center}
\end{figure}


The energy conditions as obtained are plotted in Figure 4. The energy conditions in the figures are calculated for $k=0.65$ and $\lambda=0.8$.  The present model satisfies all energy conditions except SEC. The behaviour of the energy conditions remain the same for different choices of the scaling constant $\lambda$. However, for large values of $\lambda$, say $\lambda = 20$ or more, the behaviour may change. Since we are interested in modified gravity models close to GR, we assume a small value of the scaling constant $\lambda$ and in these range of $\lambda$, the behaviour of the energy condition remains almost the same as shown in the figure.\\
\begin{figure}[t]
\begin{center}
\includegraphics[width=0.8\textwidth]{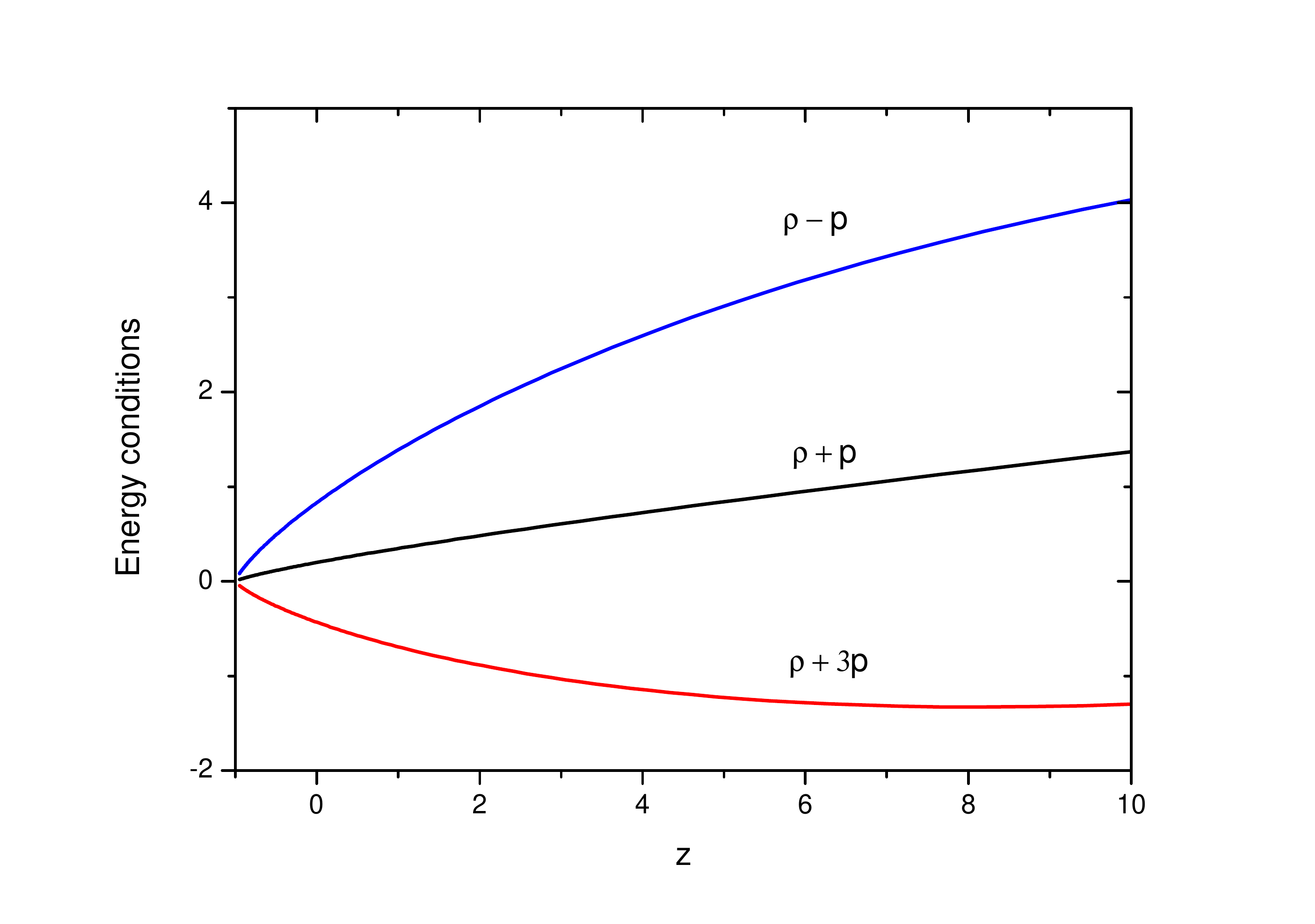}
\caption{Energy conditions for the model shown for $\lambda =0.8$ and $k=0.65$.}
\end{center}
\end{figure}

We have explored the effect of anisotropy on the dynamics of the universe as well as on the energy conditions. Here $k$ is taken as a free parameter. Anisotropy substantially affects the dynamics as is evident from Figure 5. In the figure,  $\omega$ is plotted for three values of $k$ namely $0.65, 0.8$ and 0.9. It is observed that, effect of anisotropy is more visible at early phase of evolution compared to that at late times. An increase in the value of $k$ increases the value of $\omega$ at an early time and decreases the value at late times. In other words, model with higher $k$ has less value of $\omega$ at late epoch. Interestingly all the curves with different values of anisotropy intersect at a particular redshift $z\sim 4.08$ when the EoS parameter is $-0.562$. At the present epoch, the anisotropic effect on the equation of state is better understood through its values $\omega= -0.614, -0.65$ and -0.678 corresponding to $k=0.65, 0.8, 0.9$. In order to assess the fact that, out of two parameters, $\lambda$ and $k$,  which one affects the EoS parameter to a greater extent, we have plotted the EoS parameter at the present epoch as a function of the scaling constant $\lambda$ for four different values of $k$ in Fig. 6. It is obvious from the plot that, for a given value of $k$, EoS parameter is marginally affected for a variation of $\lambda$ within a suitable range. However, with an increase in $k$, the EoS parameter decreases substantially.

\begin{figure}[t]
\begin{center}
\includegraphics[width=0.8\textwidth]{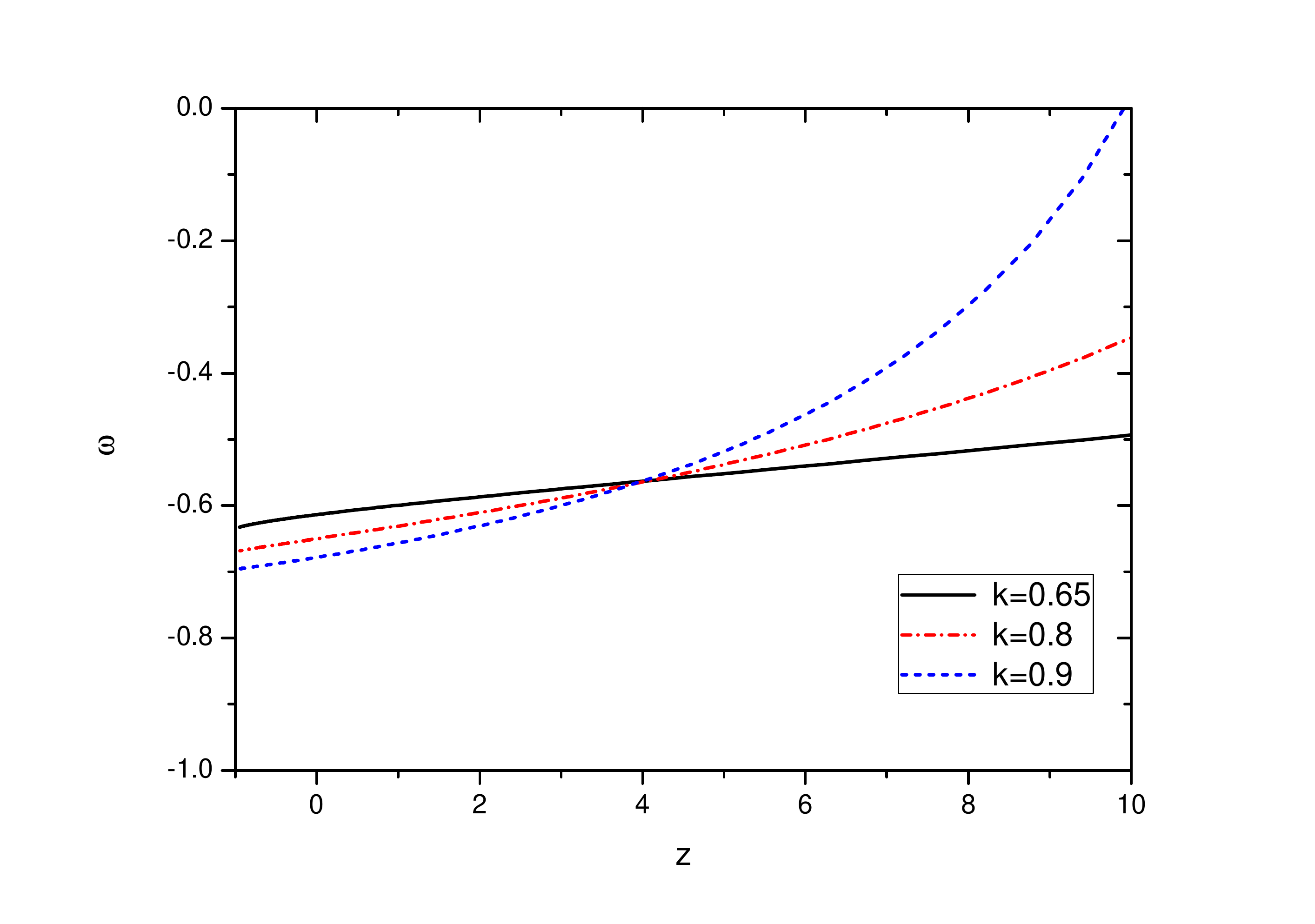}
\caption{Effect of anisotropy parameter $k$ on the EoS parameter.}
\end{center}
\end{figure}

\begin{figure}[t]
\begin{center}
\includegraphics[width=0.8\textwidth]{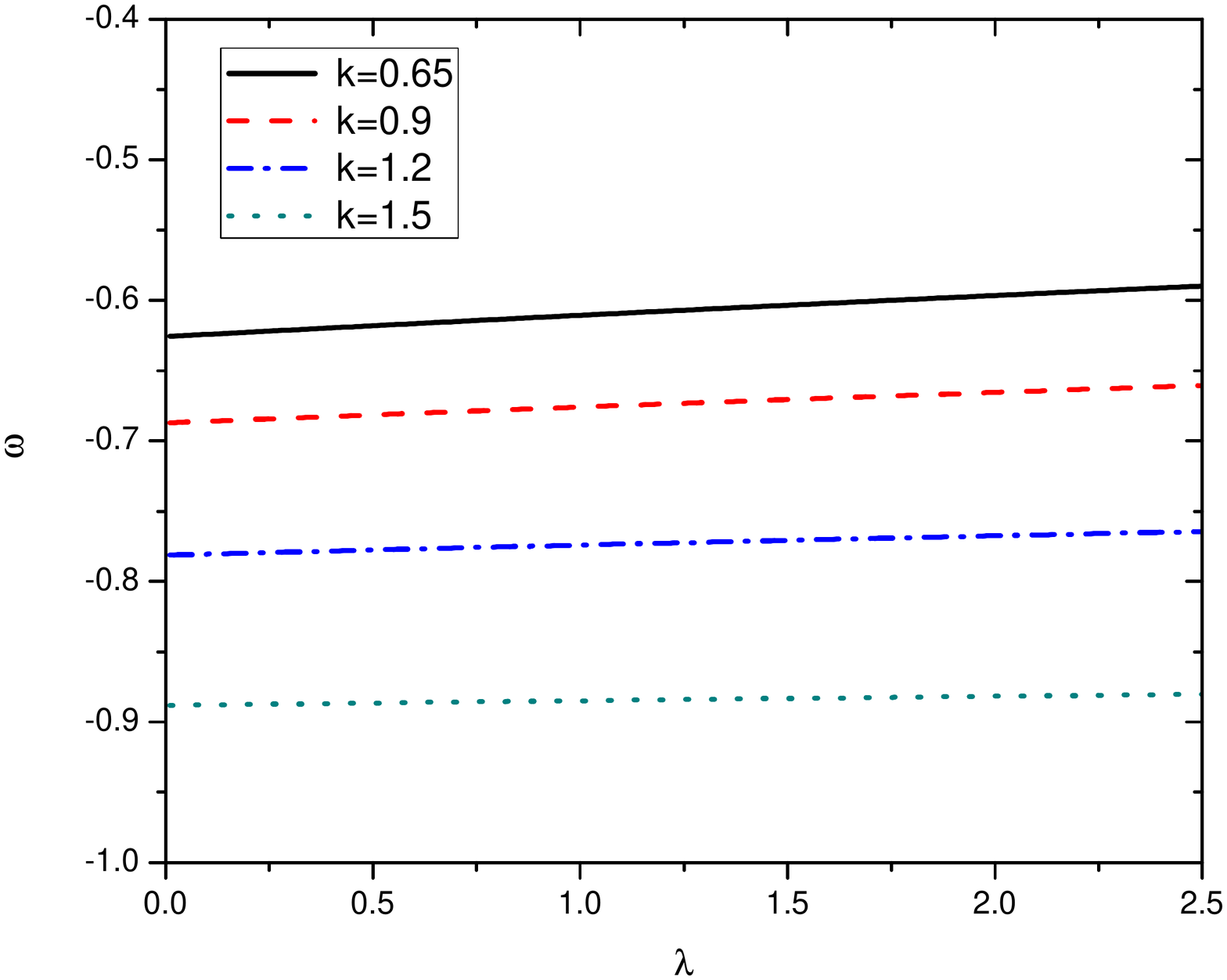}
\caption{EoS parameter $\omega$ at the present epoch as a function of scaling constant $\lambda$. The effect of anisotropy  is also shown for four representative values of $k$.}
\end{center}
\end{figure}

The energy density of anisotropic fluid $\rho_B$ considered along $x$-direction is also affected by the variation in directional anisotropy rates in the same manner as that of $\omega$. In Figure 7(a), the effect of anisotropy on the anisotropic fluid density is shown for three representative values of $k$. One can note that, in the late phase the requirement of an anisotropic fluid is much less for higher anisotropy whereas at an initial cosmic phase, more anisotropic fluid is required to maintain a higher anisotropic expansion rate. However, all the curves of the anisotropic fluid energy density intersect at a redshift $z\sim 4.08$. \\
 
\begin{figure}[ht!]
\begin{center}
\includegraphics[width=0.8\textwidth]{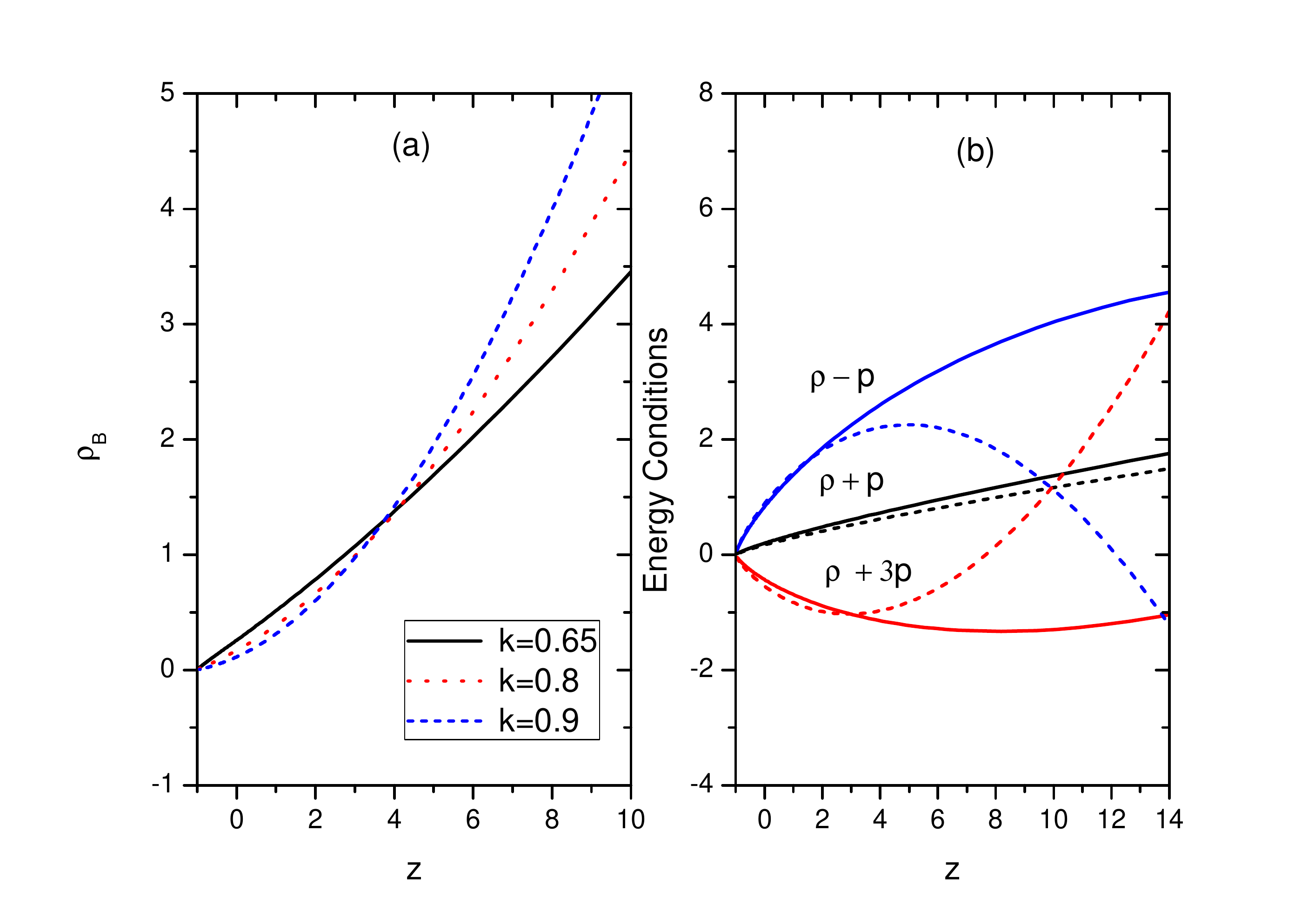}
\caption{(a) Effect of anisotropy on the anisotropic fluid density (b) Effect of anisotropy on energy conditions. Solid curves in (b)refer to $k=0.65$ and the dotted curves to $k=0.9$ for the respective energy conditions.}
\end{center}
\end{figure}

In Figure 7(b) we have shown the effect of anisotropy on the energy conditions. We have considered two representative values 0.65 and 0.9 of the anisotropic parameter $k$ for a given value of $m$ and $\lambda$. As shown in the figure, the effect of anisotropy is dramatic. Anisotropy affects marginally to NEC. An increase in the value of $k$ lowers the value of $\rho+p$ at late phase. However, at a late epoch, it is almost unaffected by the choice of $k$. Also, in case of SEC and DEC, the anisotropy has little affect during the late epoch. But at an initial epoch, an increase in the value of $k$ decreases the value of $\rho -p$ and it goes down to the negative domain violating the dominant energy condition.  With an increase in $k$, $\rho+3p$ increases into the positive domain for higher $z$ values thereby enabling the model to satisfy the strong energy condition.

The dynamical features can also be assessed from a dimensional analysis of the quantities. In eqns. \eqref{eq:18}-\eqref{eq:20}, the dynamical behaviour is governed by two terms: one depending on $t^{-2}$ and other on $t^{-\frac{2km}{k+2}}$. Since the parameters $m$ and $k$ are dimensionless, we expect that the two terms appearing in these quantities must have the same dimensions as quantified. The parameter $m$ is already constrained from the deceleration parameter. In view of the argument on the basis of qualitative dimensional analysis, we can put a constraint on $k$ as $km=k+2$. With this constraint on $m$ and $k$, pressure, energy density and energy density for the anisotropic fluid reduce to

\begin{eqnarray}
p &=& -\frac{1}{1-4\alpha^2}\left[\frac{1}{2}(3m^2-14m+3)-\alpha (m-1)^2+2(1+2\alpha)\right]\frac{1}{t^2} ,\\ \label{eq:23}
\rho &=& \frac{2}{1-4\alpha^2}\left[(m-1)\left\{(3m-7)+2\alpha (m-3)\right\}+(1+2\alpha)\right]\frac{1}{t^2} ,\\\label{eq:24}
\rho_B &=& \frac{1}{1-2\alpha}\left[m(m-1)(m-3)+4\right]\frac{1}{t^2}\label{eq:25}.
\end{eqnarray}

For a given scaling constant $\lambda$, the magnitude of these quantities now quadratically decrease with time. Consequently, the equation of state parameter becomes $\omega=-1+\frac{(9m^2+54m+33)+2\alpha(3m^2-14m+19)}{(12m^2-40m+29)+2\alpha(m^2-4m+4)}$ which is a non evolving constant and depends only on $\lambda$ and $m$. For $\lambda=0.8$ and $m$ as constrained from the deceleration  parameter, we obtain $\omega= -0.577$. With the constraint from dimensional analysis, the effective cosmological constant becomes $\Lambda = \frac{(m-1)^2}{1+2\alpha}\frac{1}{t^2}$. The evolutionary behaviour of the effective cosmological constant remains the same as before. It decreases quadratically with time in the positive domain.\\
\section*{5. Conclusions}
This paper reports the investigation of dynamical features of the cosmological models in a Bianchi type $VI_{h} (h=-1)$ space-time in presence of some anisotropic sources in $f(R,T)$ theory. It is observed that the choice of the scaling constant controlled the behaviour of EoS parameter. The effective cosmological constant is time dependent and dynamically decreases with cosmic time irrespective of the choice of the scaling constant. The energy conditions which are few additional conditions for the matter content of the gravitational theory remain same for different values of the scaling constant; however for a higher value of the scaling constant, the behaviour may change. An increase in cosmic anisotropy within the present formalism substantially affect the energy conditions. It is also observed that more anisotropic fluid is required to maintain a higher anisotropic expansion rate at an initial cosmic phase.
  
\section*{Acknowledgments} BM and SKT acknowledge the support of IUCAA, Pune, India during an academic visit where  a part of this work is done.The authors are highly grateful to the anonymous reviewers for their critical suggestions which have greatly improved the quality of the present work.

\end{document}